# Optical Properties of Organic Haze Analogues in Water-rich Exoplanet Atmospheres Observable with *JWST*

Chao He[1,2*], Michael Radke[1], Sarah E. Moran[1,3], Sarah M. Hörst[1,4], Nikole K. Lewis[5], Julianne I. Moses[6], Mark S. Marley[3], Natasha E. Batalha[7], Eliza M.-R. Kempton[8], Caroline V. Morley[9], Jeff A. Valenti[4], & Véronique Vuitton[10]

[1]Department of Earth and Planetary Sciences, Johns Hopkins University, Baltimore, MD, USA che13@jhu.edu

[2]School of Earth and Space Sciences, University of Science and Technology of China, Hefei, China

[3]Lunar and Planetary Laboratory, University of Arizona, Tucson, AZ, USA

[4]Space Telescope Science Institute, Baltimore, MD, USA

[5]Department of Astronomy and Carl Sagan Institute, Cornell University, Ithaca, NY, USA

[6]Space Science Institute, Boulder, CO, USA

[7]NASA Ames Research Center, Moffett Field, CA, USA

[8]Department of Astronomy, University of Maryland, College Park, MD, USA

[9]Department of Astronomy, the University of Texas at Austin, Austin, TX, USA

[10]Univ. Grenoble Alpes, CNRS, IPAG, 38000 Grenoble, France

Total pages: 29

The number of references: 55 (main text) +24 (Methods)

1 table and 4 figures

***JWST*** **has begun its scientific mission, which includes the atmospheric characterization of transiting exoplanets. Some of the first exoplanets to be observed by *JWST* have equilibrium temperatures below 1000 K, which is a regime where photochemical hazes are expected to form. The optical properties of these hazes, which controls how they interact with light, are critical for interpreting exoplanet observations, but relevant data are not available. Here we measure the density and optical properties of organic haze analogues generated in water-rich exoplanet atmosphere experiments. We report optical constants (0.4 to 28.6 μm) of organic haze analogues for current and future observational and modeling efforts covering the entire wavelength range of *JWST* instrumentation and a large part of *Hubble*. We use these optical constants to generate hazy model atmospheric spectra. The synthetic spectra show that differences in haze optical constants have a detectable effect on the spectra, impacting our interpretation of exoplanet observations. This study emphasizes the need to investigate the optical properties of hazes formed in different exoplanet atmospheres, and establishes a practical procedure to determine such properties.**

## 1. INTRODUCTION

Space-based and ground-based observations[1-6] indicate that many exoplanets could possess cloud and haze particles in their atmospheres and that these particles impact observed spectra. Recent modeling[7,8] and laboratory studies[9-12] suggest that organic haze particles are produced photochemically in temperate (<1000 K) exoplanet atmospheres, which are prime targets of observations for assessing habitability and searching for biosignatures beyond the Solar System. However, many exoplanets in this temperature range (including super-Earths and mini-Neptunes) have no Solar System analogs. The compositions and properties of organic hazes might be very distinct from what we know for Solar System bodies. These organic hazes can alter the transmission, emission, and reflected light spectra of exoplanets.[8,13] The optical properties of these hazes are essential to interpret exoplanet spectroscopic data and understand their atmospheres.

Because the optical properties of exoplanet organic hazes are not yet known, the optical constants of the Titan haze analogue (produced with 10% $CH_4$ in $N_2$) from Khare et al. (1984)[14] and of soots (carbonaceous particles formed from incomplete combustion of hydrocarbons)[15] are widely used to generate models and interpret observations.[8,16,17] In reality, the organic hazes formed in diverse atmospheres are expected to have a variety of compositions[18] and therefore diverse optical properties. The optical properties of Khare's Titan haze analogue or soots simply cannot represent the wide variety of atmospheric hazes we anticipate on exoplanets.[9-12] It is therefore necessary to measure the optical properties of organic haze analogues formed over a broad range of simulated atmospheric conditions, given that JWST has begun to deliver unprecedented observations of various exoplanets.[6] Exoplanets are expected to exhibit a wide diversity of atmospheric compositions. The signature of water is particularly sought on potentially habitable exoplanets because water is a key element for life on Earth. For example, water vapor has been detected in the atmosphere of a non-terrestrial exoplanet (K2-18 b) in the habitable zone.[19,20] Modeling studies[21-25] suggest that water worlds could be very common among low mass exoplanets and there might be many water-rich atmospheres. Previous laboratory studies[9,10,18] have shown that water-rich atmospheres are likely to result in organic haze formation. Opacity from organic hazes can mask spectral features from water and other gases.[2,3,26,27] Here we measure the density and the optical properties of laboratory-generated solid particles analogous to those produced in temperate water-rich atmospheres. Their optical constants (the real refractive index, *n*, and the imaginary part of the refractive index, *k*) are derived from 0.4 to 28.6 μm, covering optical wavelengths accessible with *Hubble* and ground-based facilities and the entire *JWST* wavelength range. Fig. 1 summarizes our experimental setup, the initial gas compositions and conditions, the measurements, and the analytical method of this study (for detailed information, see Methods).

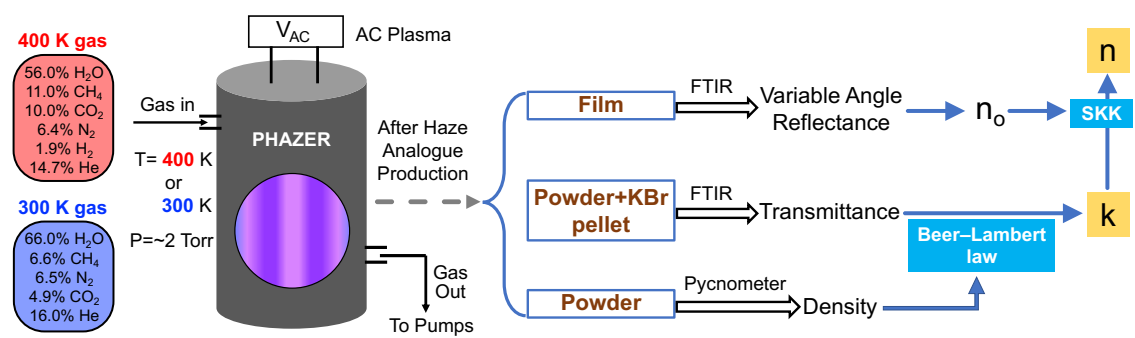

**Fig. 1.** Simplified schematic of the experimental setup, the simulated atmospheric compositions and conditions, the measurements, and the analytical method for the current study. Two haze analogues are produced with the PHAZER chamber by exposing the gas mixture (300 K or 400 K) to an AC plasma energy source. These two gas mixtures represent the equilibrium compositions for atmospheres with 1000 times solar metallicity at 300 and 400 K, which are guided from chemical-equilibrium calculations (see Methods). The high haze analogue production rate in these two gas mixtures[10] allows further analysis of the properties of the resulting particles. The density of the haze analogues is determined using a gas pycnometer, and their transmittance and reflectance spectra are measured with a Fourier transform infrared (FTIR) spectrometer. The imaginary part of the refractive indices ($k$) are calculated based on the Beer-Lambert law, and the real refractive indices ($n$) are derived using the subtractive Kramers-Kronig (SKK) relation between $n$ and $k$.

## 2. RESULTS

*2.1. Density of Exoplanet Haze Analogues*

The measured densities are 1.328 and 1.262 g cm$^{-3}$ (measurement uncertainties <1%) for the haze analogues from the simulated exoplanet atmospheres at 300 and 400 K, respectively. The density of organic haze particles is an important property that impacts aerosol microphysics processes (coagulation, transport, and sedimentation) in exoplanet atmospheres. However, prior to this study the particle density was unconstrained due to lack of observational and experimental measurements. The haze mass density is often assumed to be 1 g cm$^{-3}$ in many microphysics models[28-30], which is lower than the density we measured here. Our measurements here provide the first experimental constraints on the density of organic haze particles formed in water-rich exoplanet atmospheres, enabling realistic analysis and interpretation of observations of such exoplanets. For instance, higher haze mass density could change the vertical distribution because denser particles could more efficiently sediment to deeper layers of an atmosphere, therefore impacting the observed spectra and how we interpret them.

The particle density of the haze analogues is controlled by their chemical compositions. Because of the distinct chemical compositions, the densities of the two exoplanet haze analogues are lower than the haze analogues produced for icy bodies[31] (for Titan, Triton, and Pluto) in our Solar System using the same experimental setup. The different experimental conditions, such as the initial gas mixtures and temperatures, lead to the distinct chemical compositions of the haze analogues (see detailed comparison in Table S1). Elemental analysis has shown that the two haze analogues for exoplanets have higher oxygen (15-17% compared to <10%) but lower nitrogen (27-21% compared to ~40%) contents relative to haze analogues for Titan, Triton and Pluto.[18,32] Besides the differences in elemental compositions, the chemical structures in each haze analogue, such as their molecular weight, polarity, and degree of unsaturation, also affect their density.

*2.2. Transmittance of Exoplanet Haze Analogues and their Functional Groups*

Exoplanet atmospheric transmission observations, obtained as the planet passes in front of its host star, will be shaped by all components in the planet's atmosphere, including hazes. Fig. 2 shows the transmittance of the two exoplanet haze analogues (300 and 400 K powder+KBr pellet samples). For each sample, the transmittance spectra are obtained at three different concentrations in potassium bromide (KBr) to capture both strong and weak absorption features. As shown in Fig. 2, the general spectral shape and features are similar for the samples produced at different temperatures. Note that the spectra below 0.413 μm could be affected by the instrument as the measurements are close to the short wavelength limit of the spectrometer. The spectra are featureless from 0.413 to 2.5 μm but the absorption becomes weaker as the wavelength increases. The stronger absorptions in the shorter wavelengths (0.413-0.45 μm) suggest that the samples contain aromatic compounds and/or unsaturated species with conjugated pi bonds.[33,34] Many absorption features appear from 2.5 to 20 μm due to bond vibrations of various organic functional groups in the haze analogues.

We expand this region in Fig. 2C to show the characteristic frequencies of different bonds. Fig. 2C identifies the bonds responsible for each spectral feature, including the characteristic absorptions of O–H, N–H, C–H, C≡N, –N=C=N–, C=O, C=N, C=C, N–O, C–O, and C–N bonds.[35,36] Table 1 summarizes the vibration modes of these bonds and their peak intensities. The absorption feature at 3500–2500 cm$^{-1}$ (2.86–4.00 μm) is due to O–H bond stretching. This strong, broad feature indicates the predominance of alcohols (3500–3200 cm$^{-1}$) and carbonic acids (3300–2500 cm$^{-1}$) in the haze analogues. The stretching of N–H (3350–3300 and 3215–3190 cm$^{-1}$) and C–H (2965, 2933, 2875 cm$^{-1}$) bonds is present in this range as well but appears as relatively sharp peaks. Other absorption features in the spectra indicate the presence of nitriles, aromatics, and unsaturated and saturated organics in both the 300 and 400 K haze analogues. A small amount of carbodiimide compounds (–N=C=N–, ~2100 cm$^{-1}$) are only present in the 300 K haze analogue, probably because lower temperature promotes carbodiimide formation from cyanamide.[37] Compared to Titan haze analogues[38,39], the exoplanet haze analogues in this study have obvious compositional differences, which include various oxygen-containing groups. This is not unexpected because there are two major oxygen-containing gases ($H_2O$ and $CO_2$) in the initial gas mixtures. The presence of oxygen-containing groups in the exoplanet haze analogues is supported by other analyses. Elemental analysis shows that these two haze analogues have large oxygen contents (15% or 17% in mass), and high-resolution mass spectra reveal many oxygen-containing molecules in these samples.[18] The compositional difference leads to distinctive optical properties of the exoplanet haze analogues.

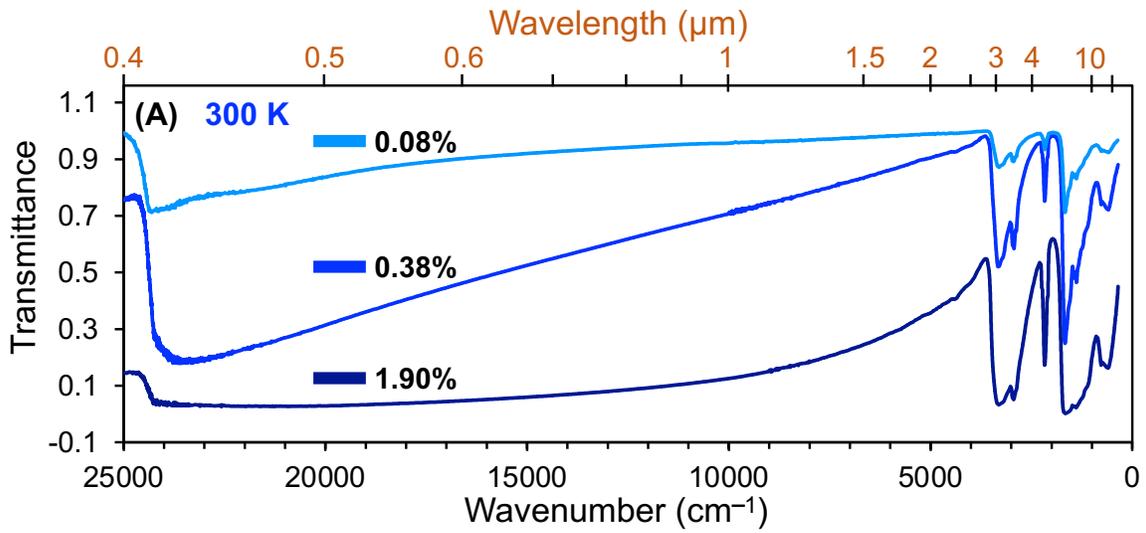
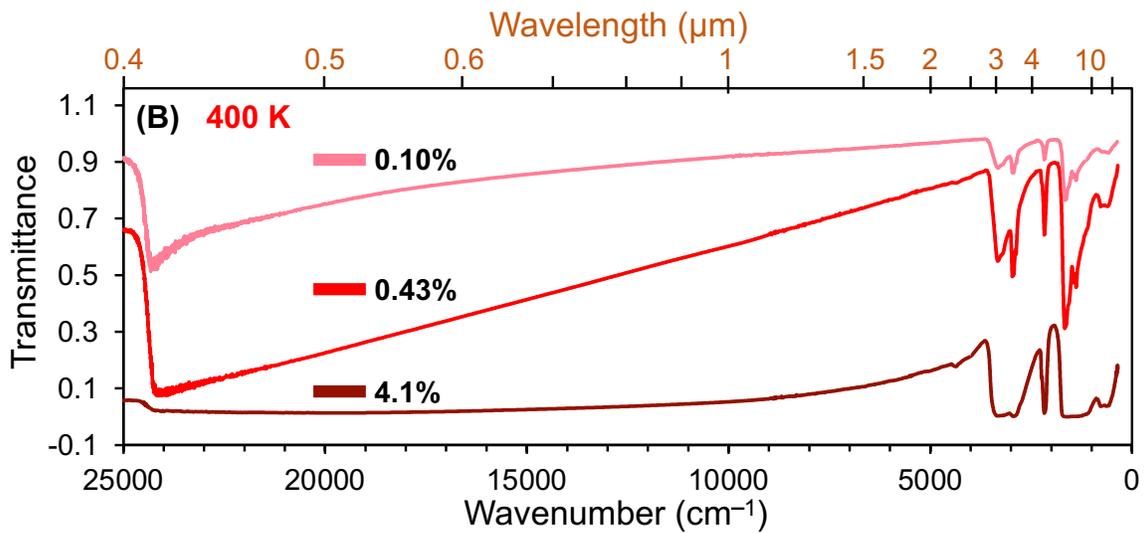
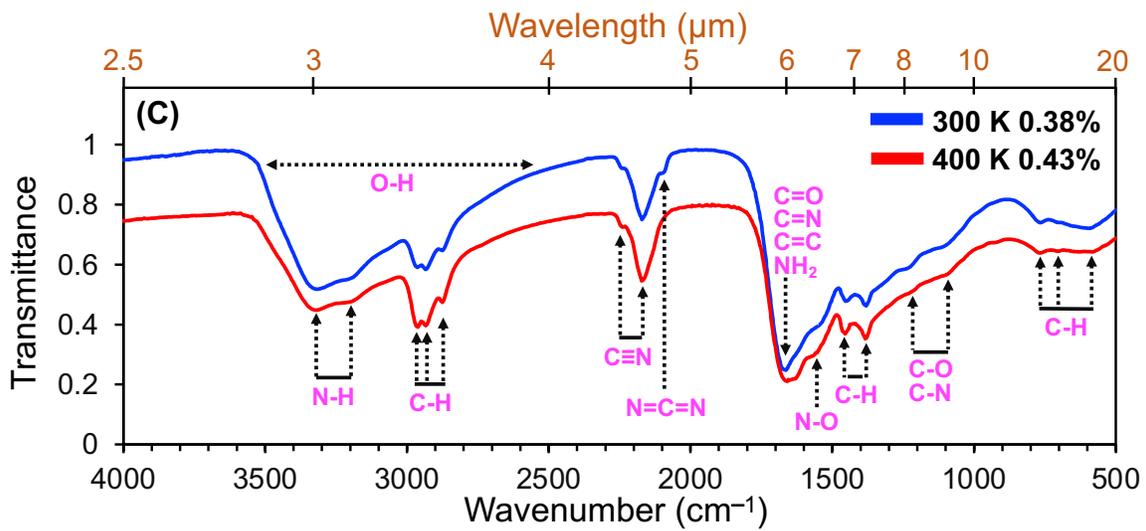

**Fig. 2.** The transmittance spectra of two exoplanet haze analogues formed in water-rich gas mixtures at 300 K (A) and 400 K (B). The lines with different shades of color show the spectra of the samples with different concentrations in KBr (lighter color indicates lower concentration and darker indicates higher). The samples are mixed with KBr, and the percentage concentration is by mass. The transmittance spectra are obtained at different concentrations to capture both strong and weak absorption features in the samples. Panel C is the expansion of two spectra from panels A (0.38% of the 300 K sample) and B (0.43% of the 400 K sample) from 2.5 to 20 μm to show the absorption features of various functional groups. Different bonds are labeled in the figure near their absorption features.

**Table 1.** Functional group assignments to characteristic absorption frequencies of two exoplanet haze analogues as observed in the transmittance spectra (Fig. 2)

| Frequency (ν) cm$^{-1}$ | Wavelength (λ) μm | Functional group | Intensity |
|---|---|---|---|
| 3500-2500 | 2.86-4.00 | -O-H stretching, in alcohols and carbonic acids | strong, broad |
| 3350-3300 | 2.98-3.03 | -NH$_2$ or -NH- stretching | strong |
| 3215-3190 | 3.11-3.13 | -NH$_2$ stretching or overtone of NH$_2$ bending | strong |
| 2965 | 3.37 | -CH$_3$ asymmetric stretching | strong, sharp |
| 2933 | 3.41 | -CH$_2$- asymmetric stretching | strong, sharp |
| 2875 | 3.48 | -CH$_3$ symmetric stretching | strong, sharp |
| 2244-2232 | 4.46-4.48 | -C≡N or R-C≡C-R stretching | medium, shoulder |
| 2172 | 4.60 | Conjugated -C≡N stretching | strong, sharp |
| 2108-2094 | 4.74-4.77 | -N=C=N- stretching | medium, shoulder |
| 1680-1626 | 5.95-6.15 | C=O, C=N, C=C stretching or -NH$_2$ scissors bending | very strong |
| 1580-1540 | 6.33-6.49 | C=C stretching (aromatic) or N-O stretching (nitro) | medium, shoulder |
| 1462-1450 | 6.84-6.90 | sp$^3$ C-H bending | strong, sharp |
| 1386-1378 | 7.21-7.26 | sp$^3$ C-H bending or aldehydic C-H bending | strong, sharp |
| 1247-1224 | 8.02-8.17 | C-O or C-N stretching (aromatic) | weak, shoulder |
| 1126-1104 | 8.88-9.06 | C-O or C-N stretching (aliphatic) | weak, shoulder |
| 775-762 | 12.9-13.12 | sp$^2$ C-H bending | medium |
| 714-701 | 14.0-14.26 | sp$^2$ C-H bending | medium |
| 612-584 | 16.34-17.12 | sp$^2$ C-H bending | medium |

*2.3. Optical Constants of Exoplanet Haze Analogues*

Optical constants for haze analogues can be used to incorporate their features into theoretical spectra and for comparisons with optical constants of exoplanet haze retrieved from observations. Fig. 3 shows the optical constants of two exoplanet haze analogues (300 and 400 K). The real refractive indices (*n*) are in the range between 1.43 and 1.82 for both samples, and the imaginary part of the refractive indices (*k*) vary in a broad range from ~$10^{-3}$ to $10^{-1}$. The determined *n* and *k* values and their uncertainties can be found in Supplementary Information (Table S2). The uncertainties of the *k* values are estimated by considering the uncertainties of the transmittance measurements, the thickness calculations, and the average relative standard deviation of *k* values obtained for the samples at three different concentrations. For the *n* values, the uncertainties are determined from the error propagations of the uncertainty of $n_0$ at the anchor point and the integration of the *k* values. The uncertainties for the *k* values are about 1-4% at longer wavelengths (>2.9 μm) where the *k* values are larger than 0.01, but increase up to 9% at shorter wavelengths (0.4 to 2.9 μm) where the *k* values are smaller than 0.01. The determined *n* values have small uncertainties (3-4%) across the measured wavelength range because of the accuracy of the $n_0$ (uncertainty less than 1%) and the integration of the *k* values over the entire wavelength range. The *n* and *k* values of the two haze analogues (300 and 400 K) are similar in general and share many fine-scale features. Over most of the measured wavelength range, the 300 K sample has slightly lower *n* values but slightly higher *k* values relative to the 400 K sample. This could be caused by differences in their chemical compositions. For example, the 300 K sample has higher *k* values at ~3 μm, indicating that it contains more N–H bonds. This is consistent with the elemental analysis that shows the 300 K sample has higher nitrogen content than the 400 K sample (27% vs 21%).[18] In addition, the higher *k* values at 5.9-6.5 μm suggest that there are more double bonds in the 300 K sample.

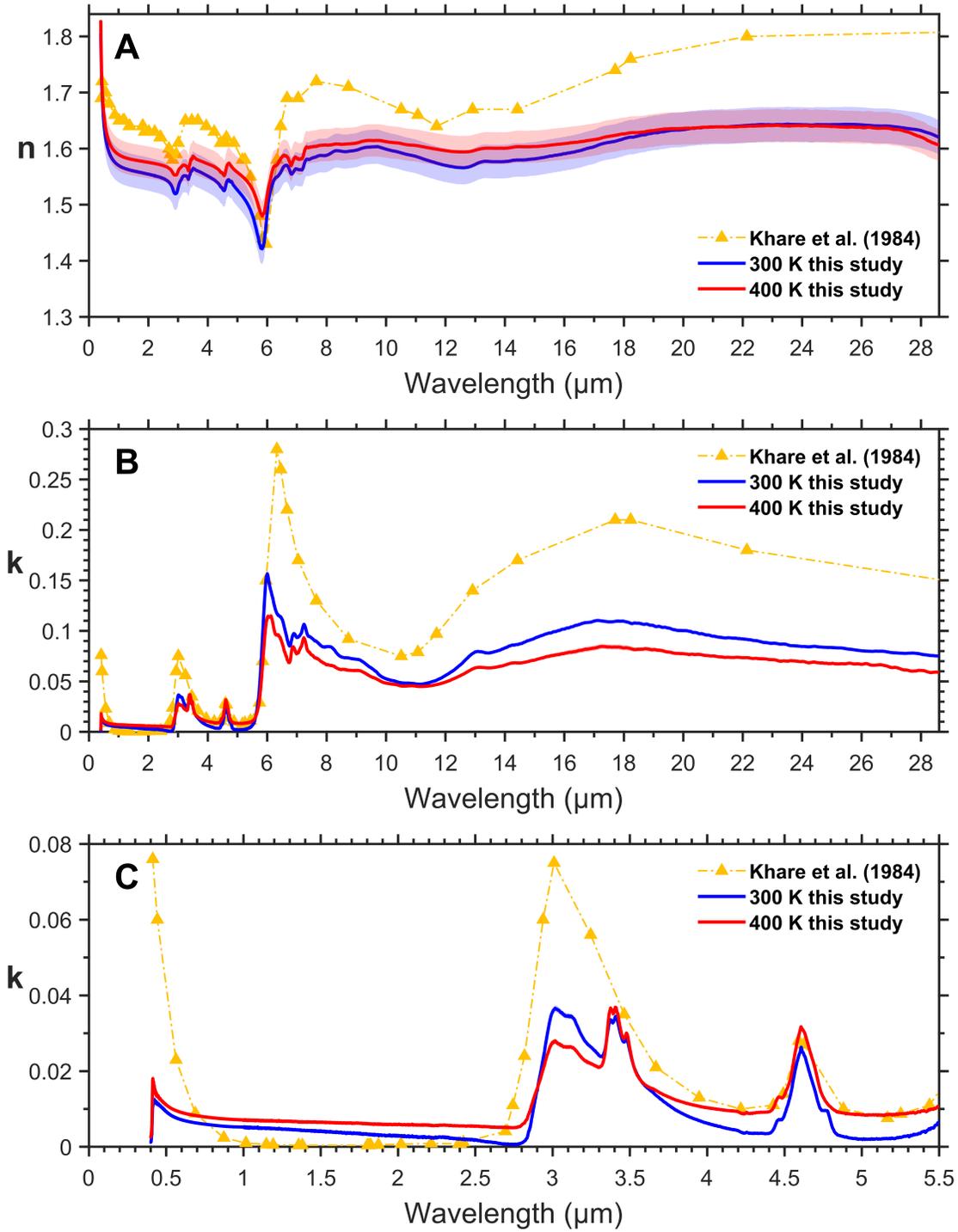

**Fig. 3.** Optical constants of two exoplanet haze analogues along with those of the Titan haze analogue from Khare et al (1984): (A) the real refractive indices (*n*), and (B) the imaginary part of the refractive indices (*k*) as function of wavelength (0.4 to 28.6 μm). Panel C shows the difference of *k* values between this study and the one from Khare et al

(1984), when *k* is small in the short wavelengths (0.4 to 5.5 µm). The shaded area shows the uncertainties of our derived *n* and *k* values. Note that the data below 0.413 µm could be affected by the instrument as the spectral measurements in this region are close to the short wavelength limit of the spectrometer, and should be used with caution.

The optical constants of the Titan haze analogue from Khare et al. (1984)[14] are also plotted in Fig. 3 for comparison. Due to their extensive spectral coverage (from soft X-ray, 0.025 µm, to microwave frequencies, 1000 µm), the optical constants from Khare et al. (1984)[14] are widely used in atmospheric modeling and observation interpretations for both Solar System bodies (Titan[40], Triton[41], and Pluto[42]) and many exoplanets.[16,17,29,30,43] However, the haze analogues used in Khare et al. (1984)[14] for the optical measurements were produced with DC plasma discharge in $N_2/CH_4$ = 90/10 gas mixture at 0.2 mbar, with the purpose of simulating haze particles formed in Titan's upper atmosphere.

As shown in Fig. 3, the general trend of the optical constants of the two exoplanet haze analogues in this study is similar to that of the Titan haze analogue from Khare et al. (1984)[14] because of the organic nature of these materials. However, our *n* and *k* values are substantially smaller than those of Khare et al. (1984)[14] over most of the measured wavelength range. Both of our samples have smaller n values than the Titan haze analogue from Khare et al. (1984)[14] except at 5.8-6.2 µm where the n values of our 300K sample are similar to those of Khare et al. (1984).[14] The *k* values from Khare et al. (1984)[14] at 0.4-0.6 µm, ~3 µm, and above 6 µm are about two times greater than our *k* values; they are comparable to the *k* values of our 400 K sample from 3.4 to 5.8 µm; but they are smaller than our *k* values in the short wavelength region (0.7-2.6 µm). Besides the differences of the absolute *n* and *k* values, the data from Khare et al. (1984)[14] has fewer features than our values mainly due to two reasons. First, the spectral measurements in Khare et al. (1984)[14] were done at very low resolution, especially over the longer wavelengths (6-28.6 µm); second, the compositions are distinct between our samples and those in Khare et al. (1984). The low-resolution can lead to fine features not being resolvable (such as those from 6-10 µm), and the compositional difference may cause some features to be weaker or absent (such as those at 3-3.5 µm). The higher spectral resolution of the current study reveals new spectral features that may enable measurements of haze composition, rather than just haze detection. Nonetheless, the differences in the absolute values and the features can have consequential impacts on atmospheric modeling and observational interpretation of exoplanets.

*2.4. Effect of New Optical Properties of Exoplanetary Hazes on Atmospheric Transmission Spectra*

We next briefly explore the effects of these newly derived exoplanet haze analogue optical properties on transmission spectra for a representative exoplanet atmosphere. Since the experiments were run for hazy, water-rich atmospheres, we choose the parameters of the well-studied, likely aerosol-laden exoplanet GJ 1214 b as a reasonable example planet to demonstrate how our new optical properties would affect observations. GJ 1214 b has a mass of $8.17 \pm 0.43$ $M_\oplus$, a radius of $2.742 \pm 0.05$ $R_\oplus$, and an equilibrium temperature of 596 K.[44] The planet's mass and radius suggest its bulk density could be consistent with a hydrogen-rich atmosphere, or a steam atmosphere with a significant hydrogen/helium fraction[44] as in our experiments. The temperature is sufficiently warm that water should not condense out of the upper atmosphere. Observations with *Hubble* show its atmosphere has a featureless transmission spectrum from 1.1 to 1.7 microns, which could be indicative of a high altitude (~mbar) haze layer.[1] The incident UV flux is expected to drive significant photochemistry and potentially haze formation.[45,46]

Fig. 4 shows the synthetic spectra of a water-rich atmosphere with the initial gas composition from the experiments at 400 K around a GJ 1214 b-like planet with the effects of a haze layer. We use the open-source aerosol modeling code *Virga*[47,48] and the open-source radiative transfer suite *PICASO*[49] to compute Mie properties for our haze analogues and implement them into a synthetic transmission spectrum model of an exoplanetary atmosphere (for detailed information about the haze modeling and generating the synthetic spectra, see Methods). We show spectra based on our newly derived optical properties for 300 and 400 K water-rich atmospheric haze analogues and for the commonly used Titan haze analogue of Khare et al. (1984)[14]. Our aim in this work was to isolate the effect of the different haze analogue optical properties on the simulated atmospheric transmission spectra rather than perform a full model parameter study, and so we used fixed values for particle size distribution, number density, vertical mixing (see Table S3, those values are either consistent with lab measurements or typical values used to model GJ 1214 b). Interpretation of actual atmospheric spectral data would require a full investigation into all these variables but is beyond the scope of this work.

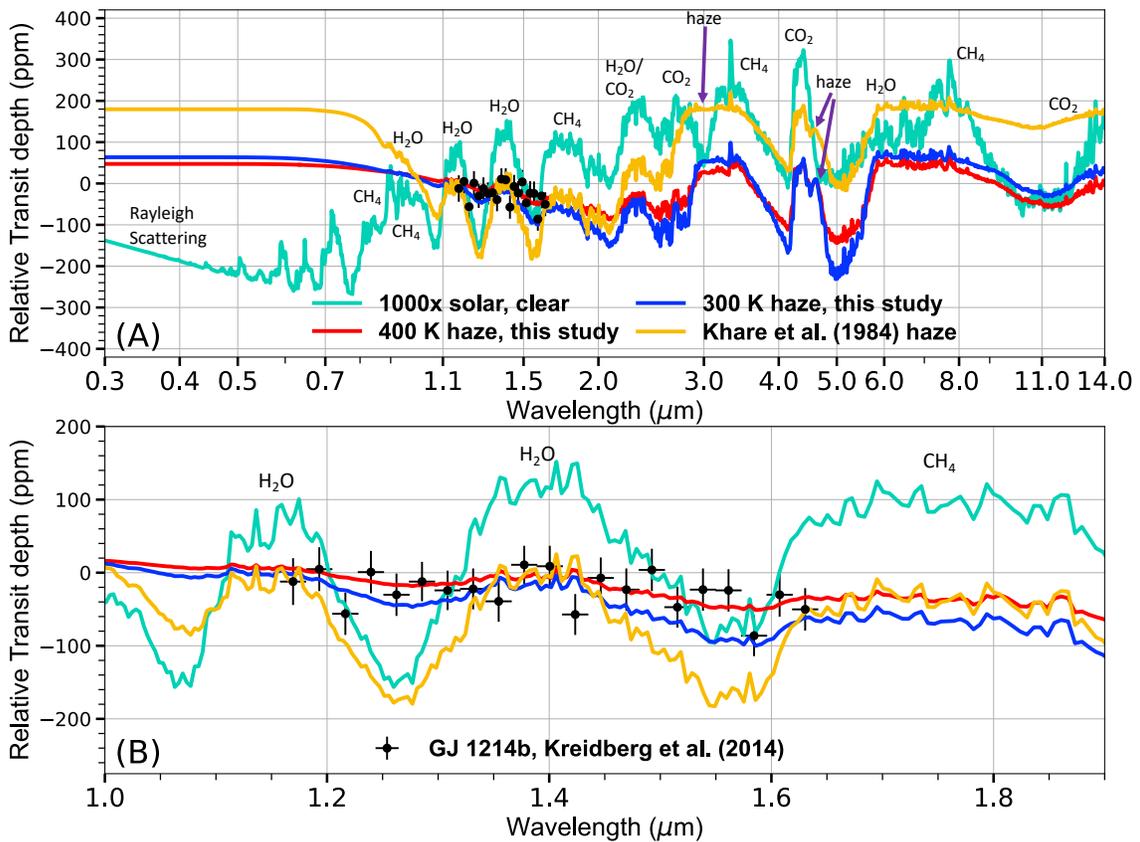

**Fig. 4.** Model spectra of a water-rich atmosphere around a GJ 1214 b -like planet, showing the effect of our newly measured haze analogue optical properties. We also show the existing *Hubble* data of GJ 1214 b (Kreidberg et al. 2014)[1]. The top panel (A) shows the spectra of our modeled atmospheres from 0.4 to 14 microns; bottom panel (B) shows our modeled atmospheres focused on *Hubble* Wide Field Camera 3/G141 instrument wavelengths. In teal, we show models of clear atmospheres with the 400 K laboratory atmospheric composition. Orange line shows a modeled hazy atmosphere using the optical properties for Titan haze analogue as measured by Khare et al. (1984). The red and blue lines show models generated using our newly derived optical properties for laboratory-generated haze analogues at 400 and 300 K, respectively. The molecules and/or haze analogues responsible for the simulated atmospheric features are indicated in each plot. **Given the same other atmospheric assumptions, our newly measured haze analogues affect the simulated spectra differently than the Titan haze analogue, and such differences would be detectable by existing space-based observatories like *Hubble* and *JWST*, as with the GJ 1214 b data shown here.**

Throughout the spectrum, the values of $n$ and $k$ between our newly measured water-rich atmospheric haze analogues and those of the Titan haze analogue[14] show similar behavior but at different intensities, as shown in Fig. 3. As such, all the hazy synthetic spectra we show in Fig. 4 mute the gaseous spectral features of the atmosphere, but to differing extents depending on the optical properties assumed for the haze particles. Notably, the Khare Titan haze analogue more heavily mutes the atmospheric spectral features longward of 3.0 µm, while our water-rich atmospheric haze analogues more heavily mute the atmospheric spectrum between 0.7 and 2.5 µm– that is, the wavelength region of focus for most *Hubble* exoplanet studies. In particular, the Titan haze analogue more strongly absorbs at ~3 µm and ~6 µm, which probe functional groups due to amine (N-H) as well as carbon- and nitrogen-bearing (C=C, C=N, $NH_2$) bonds. The transit depth differences between haze models range from ~100 ppm in the mid-IR to ~150 ppm in the visible and near-IR, all of which are well within the sensitivity of *JWST* and *Hubble* measurements.[1,51-55]

Additionally, the strength of the scattering slope from the visible towards the near-UV is enhanced by all haze analogues, though the Titan haze analogue absorbs more strongly because of its larger $k$ over the shortest wavelengths (0.4 to 0.7 µm). The difference between the visible-UV slopes of Titan haze analogue and water-rich atmospheric haze analogues is ~150 ppm for our modeled atmospheres. Smaller particles would enhance the variation between the visible-UV slopes (see Extended Data Fig. 3). Such differences are within the precision of Hubble's WFC3/UVIS instrument, which covers 0.2 to 0.8 μm. Since we have explicitly modeled the same haze particle size distributions and number densities for each set of optical properties, these differences here stem purely from the stronger absorption by the Titan haze analogue over these wavelengths. In other words, using the optical properties from this study or from Khare et al. (1984)[14] to fit spectra from real observations would result in different inferred particle sizes and number densities for the same atmosphere to achieve the same scattering slope.

Most interestingly, our water-rich atmospheric haze analogues almost completely mute the atmospheric spectrum over the HST WFC3/G141 bandpass, but the Titan haze analogue of Khare et al. (1984)[14] allows the water and methane atmospheric features to peek through from 1 to 2 μm because its $k$ values are an order of magnitude lower in the NIR. This results in our water-rich atmospheric haze analogues providing a better match to the existing GJ 1214 b data from *Hubble*[1] compared to the Khare et al. (1984)[14] Titan haze analogue for the same abundance of haze mass loading. All three sets of haze analogue optical properties allow atmospheric features to peek through at redder wavelengths which are covered by *JWST*. The size of these atmospheric features at longer wavelengths, even muted by haze, is still well within the precision of *JWST* measurements. Future model development and modeling studies, beyond the scope of this work, are needed to fully assess how vertical mixing and sedimentation of haze layers with these water-rich atmospheric haze analogue optical properties influence atmospheric spectra.

3. DISCUSSION

So far, the optical constants of the Titan haze analogue by Khare et al. (1984)[14] and various soots[15] have served as the only source for simulating the optical properties of organic haze in exoplanet atmospheres, until a recent study by Corrales et al. (2023)[56] who reported the optical constants (0.3-10 μm) of haze analogues produced from nitrogen-rich (90% or 95%) gas mixtures ($N_2$-$CH_4$-$CO_2$) and applied them to model the spectra of GJ 1214 b. Our study provides optical constants (0.4 to 28.6 μm) of laboratory-generated organic haze analogues to the haze particles expected to form in water-rich exoplanet atmospheres. The optical constants of our water-rich atmospheric haze analogues differ from those of Khare's Titan haze analogue[14], therefore affecting transmission, thermal emission, and reflected light spectra of exoplanets to different extents. When interpreting actual observations of exoplanets, using different optical constants to simulate the hazes can lead to different conclusions; in some cases, using non-representative optical constants may lead to misinterpretations. Such misinterpretations could include incorrect atmospheric abundances of key species such as water and methane[57,58], or incorrect aerosol particle sizes and abundances[30], which could then further affect the temperature structure[17] and thus the inferred dynamics and climate feedbacks within an atmosphere.[59]

JWST has an unprecedented capability to detect faint chemical signatures in exoplanet atmospheres and is set to observe a wide variety of exoplanet types. The haze particles formed in those exoplanet atmospheres are likely to have different compositions, which would then have distinct optical properties. The optical properties of different haze analogues are essential to interpreting observations, but we should be cautious when applying the existing laboratory-generated haze analogue optical constants to different types of exoplanets. In addition, the optical constant data of different haze analogues are necessary input parameters for atmospheric modeling of various exoplanets to understand their physical and chemical processes (i.e., their temperature structures, climates, etc.). The two sets of optical constants reported here are applicable for temperate water-rich exoplanet atmospheres and can be used for current and future observational and modeling efforts of such atmospheres. More laboratory work is needed to determine optical constants of haze analogues formed in various exoplanet atmospheric regimes. This study demonstrates a feasible avenue to determine such properties in the future.

## 4. METHODS

### *4.1 Production of Haze Analogues*

The exoplanet haze analogues are produced using the Planetary Haze Research (PHAZER) experimental setup at Johns Hopkins University.[31] The detailed experimental procedure has been described previously.[9-12] This study uses haze analogues produced in the two water-rich gas mixtures from our previous investigations.[9,10,18] The initial gas mixtures for our experiments are guided by the calculations from the chemical-equilibrium models of Moses et al. (2013).[60] Photochemical hazes are expected to play an important role in the atmospheres of exoplanets with equilibrium temperature below 1000 K, especially those exoplanet atmospheres with enhanced metallicities or enhanced C/O ratios.[7,9,10] The detailed compositions of actual exoplanet atmospheres have not yet been measured well, so we resort to chemical equilibrium calculations[60] to guide our experiments. The equilibrium atmospheric compositions provide a reasonable starting point to investigate the photochemical processes in the atmospheres of super-Earths and mini-Neptunes with higher metallicity. In our previous studies[9-12], we explored gas mixtures representative of atmospheres with a range of metallicities (100×, 1000×, and 10,000× solar metallicity) and equilibrium temperatures (300, 400, 600, and 800 K). The two water-rich gas mixtures represent the equilibrium compositions for atmospheres with 1000× solar metallicity at 300 and 400 K. These two gas mixtures produce haze analogues at a higher rate than the other cases[10], which allows further analysis of the properties of the resulting haze analogues. We are exploring different techniques to obtain the optical properties of the haze analogues produced in the lower production rate experiments[10] and will publish them in future work.

Here, we briefly recount the production procedure. Gas mixtures, excluding water vapor, are premixed in a stainless-steel cylinder with high-purity gases purchased from Airgas ($H_2$-99.9999%, He-99.9995%, $N_2$-99.9997%, $CH_4$-99.999%, $CO_2$-99.999%). The premixed gas mixture flows at a proportional rate of 10 standard cubic centimeters per minute (sccm) based on the mixing ratio (4.4 and 3.4 sccm for 400 and 300 K experiments, respectively). Water vapor is introduced to the system at a pressure corresponding to the mixing ratio (1.06 Torr for 400 K and 1.25 Torr for 300 K experiment), which is provided from HPLC water (Fisher Chemical) at the desired temperature maintained by a dry ice/methanol/water cold bath. A heating coil is employed to heat the gas mixture (including water vapor) to the experimental temperature (400 or 300 K). The gas mixture is then exposed to an AC glow discharge in a reaction chamber, which initiates complex chemical processes and leads to the formation of new gas-phase products and solid particles. After 72 hours of continuous discharge flow, we collect the haze analogues as powders and films on quartz substrate disks in a dry (<0.1 ppm $H_2O$), oxygen free (<0.1 ppm $O_2$), $N_2$ glove box (Inert Technology Inc., I-lab 2GB). The haze analogues are kept in the glovebox until further analysis to avoid contamination from Earth's atmosphere and light sources. Note that the energy density of the plasma source used in our experiments[11,12] is about 170 W m$^{-2}$, which is higher than the UV flux received by temperate exoplanets orbiting low mass M-star, such as GJ 1214 b (~3.5 W m$^{-2}$ for 10 to 400 nm UV photons that are important for atmospheric photochemistry). Laboratory simulations usually employ a higher energy density than those in real atmospheres to accelerate the chemistry and observe its impact within a reasonable timeframe.

Laboratory-generated haze analogues have been characterized for their chemical, physical, material, and optical properties. The optical constants of several haze analogues have been reported (e.g., Khare et al. 1984;[14] Ramirez et al. 2002;[61] Tran et al. 2003[62]; Vuitton et al. 2009;[63] Imanaka et al. 2012;[64] Sciamma-O'Brien et al. 2012;[65] Mahjoub et al. 2014;[66]

Gavilan et al. 2018;[67] Jovanovic et al. 2021;[68] He et al. 2022;[69] and Corrales et al. 2023[56]). These haze analogues were prepared using different setups and under different conditions; their optical constants were determined with different wavelength coverages using different techniques. Brassé et al.[70] reviewed the optical constants of various Titan haze analogues (produced from $N_2$-$CH_4$ gas mixtures) in 2015. Recently, Jovanovic et al. (2021)[68] reported the optical constants of Pluto haze analogues (produced from $N_2$-$CH_4$-CO gas mixtures), while Gavilan et al. (2018)[67] and Corrales et al. (2023)[56] derived the optical constants for haze analogues (produced from $N_2$-$CH_4$-$CO_2$ gas mixtures) in anoxic and oxic Earth-like exoplanet atmospheres. The optical constants vary from sample to sample but their general trends are similar, and they share the same absorption features like those at 3.4, 4.5, and 6.2 μm. The haze analogues in these previous studies are all produced from nitrogen-rich gas mixtures ($N_2 \geq 90\%$). Our study here focuses on the optical properties of haze analogues produced from water-rich gas mixtures, which have not been reported previously. For simplicity, we compare our new results to the optical constants from Khare et al. (1984)[14] because the haze analogue from Khare et al. (1984)[14] is a good representative for many other haze analogues and the optical constant data from Khare et al. (1984)[14] is used the most in exoplanet studies.[16,17,29,30,43] In addition, the wide spectral coverage (0.025-1000 μm) of their data[14] overlaps with all our measured wavelengths.

## *4.2 Density Measurements*

The density of exoplanet haze analogues has not been reported previously. We determine the density of our haze analogues by measuring the mass and volume of a certain amount of sample by using an analytical balance (±0.0001 gram) and a gas pycnometer (AccPyc II 1340, Micrometrics), respectively. We first fill the powder sample to ∼70% of a 0.1 cm³ cup and measure the mass of the sample in the cup on the balance (where the mass of the cup is known). Next, we measure the volume of the sample at ambient temperature with the gas pycnometer that uses the gas (helium) displacement method to determine the sample volume based on Boyle's law of volume–pressure relationships. We set the pycnometer to measure 20 cycles, and the average volume from the 20 measurements is reported with a standard deviation less than 0.0001 cm$^3$. The density is then calculated by dividing the mass by the volume.

*4.3 Vacuum Fourier-transform Infrared Spectroscopy (FTIR) Measurement*

We employ a Vertex 70v FTIR spectrometer (Bruker Optics) to characterize the spectral properties of the haze analogues. The Vertex 70v is a vacuum spectrometer, which can eliminate the spectral features ($H_2O$ or $CO_2$ absorption) from Earth's atmosphere and increase the peak sensitivity without masking very weak spectral features. The wavelength range of the spectrometer is 0.4 to 28.6 μm (25,000 cm to 350 cm$^{-1}$) with a maximum resolution of 0.4 cm$^{-1}$, covering the whole observable wavelength range of *JWST* and a large part of *Hubble*. This spectrometer is configured with a Reflection/Transmission accessory (A510 Q/T, Bruker) and Seagull Variable Angle Reflection Accessory (Harrick Scientific), allowing transmittance measurements and reflectance measurements at different angles.

To derive the optical constants of the haze analogues, we measure the reflectance of the film samples on quartz discs and transmittance of the powder samples using the potassium bromide (KBr) pellet method. The films deposited on optical-grade quartz substrates in the reaction chamber are directly used for the reflectance measurements (4.3.1). The procedure for making KBr pellets is described in 4.3.2 and the transmittance measurements of the pellets are described in 4.3.3.

**4.3.1 Reflectance Measurements of the films**

Using the Seagull Variable Angle Reflection Accessory, we measure the reflectance of each haze analogue film at two different angles of incidence, 15° and 45°. The reflectance of the films on quartz substrates is measured under vacuum (below 0.2 mbar) at room temperature (294 K). We measure the reflectance in the wavelength range from 0.4 to 1.1 µm using a quartz beamsplitter and silicon diode detector. We acquire 1000 scans and average them to obtain spectra with a resolution of ~5 cm$^{-1}$. The reflectance of an aluminum standard mirror is measured as reference. The reflectance spectrum of the sample is the ratio of the sample measurement to the reference measurement. Note that the chemical compositions and the thickness of the films have major impacts on the measured reflectance. The particles produced in different experiments have different morphology and size, which also affect the reflectance spectra.

**4.3.2 Preparation of the KBr Pellets**

A quality KBr pellet is critical for quantitative measurements. KBr is a typical carrier (as pellet or disk) for spectroscopy measurements because it is optically transparent from ultraviolet (0.22 µm) to far IR (~30 µm). To ensure that the pellet is dry, KBr powder is dried in an oven overnight at 110 °C and immediately transferred to the dry $N_2$ glove box after heating. Subsequent pellet-making steps are all performed in the glove box. First, the KBr and the sample powders collected in the chamber are separately ground to fine particles with ShakIR (a ball-mill grinder from Pike Technologies). Dry and fine KBr is used to make a pure KBr pellet as a spectral reference. The fine particles of the sample and KBr are mixed using the ShakIR to create a homogenous mixture of sample/KBr before making pellets.

The concentration of the sample in KBr needs to be optimized for transmittance measurements, usually between 0.5-3% depending on the absorption coefficients. Due to the complex nature of the haze analogues, we prepare the haze analogue/KBr mixtures with different concentrations. We choose three concentrations that reveal both strong and weak absorption in each sample. The actual concentrations of our mixtures range from 0.08 to 4.1%. Because the quantities of the haze analogues in the KBr pellet are relatively small, we use a two-step dilution method to minimize errors in the measurement process. As reported in Myers et al. (2019)[71], this method can reduce gravimetric error for the mass measurements from nearly 10% to less than 1%. For example, to prepare a 0.1% mixture, we weigh 25 mg of the haze analogue using the analytical balance (±0.0001 gram) and mix it with 975 mg of KBr in the ShakIR to create a 2.5% mixture; next, we take 40 mg of the 2.5% mixture and 960 mg of pure KBr, and then mix them thoroughly using the ShakIR. After two dilutions, a 1:1000 (or 0.1%) homogenous sample/KBr mixture is ready to be pressed into a pellet.

Approximately 200 mg of finely ground KBr powder or homogenous mixture is placed into a 13 mm pellet die (Pike Technologies), and then is pressed on a CrushIR 15 Ton Digital Press (Pike Technologies). We apply 3 tons of force for 2 seconds, 7 tons for 30 seconds, and 10 tons for 120 seconds. The pressed pellet (diameter: 13 mm, thickness: ~0.5 mm) is removed from the die, mounted on a standard sample holder, and loaded into the FTIR spectrometer for transmittance measurement.

**4.3.3 Transmittance Measurements of the Pellets**

We measure the transmittance of the pressed pellets over a wavelength range from 0.4 to 28.6 μm under vacuum (below 0.2 mbar) at room temperature (294 K). Two detectors (silicon diode and DLaTGS detector) and two beamsplitters (quartz beamsplitter and KBr beamsplitter) are used to cover the whole wavelength range. The silicon diode detector and quartz beamsplitter are used from 0.4 to 1.11 μm (25000 to 9000 cm$^{-1}$); the DLaTGS detector and quartz beamsplitter are used from 0.83 to 1.25 μm (12000 to 8000 cm$^{-1}$); and the DLaTGS detector and KBr beamsplitter are used from 1.11 to 28.6 μm (9000 to 350 cm$^{-1}$). Overlapping data confirm that the spectrometer is calibrated properly across different wavelength ranges. For each measurement, 500 scans are acquired with a resolution of ~2 cm$^{-1}$. The transmittance of a pure KBr pellet is measured as a reference.

The transmittance spectrum of the sample is the ratio of the sample measurement to the reference measurement. The transmittance of the pure KBr and the 300 K haze analogue in KBr is shown in Figure S1. As shown in Figure S1, the transmittance of the pure KBr pellet is above 80% across the wavelength range we measured, illustrating the low-absorbing nature of KBr. The particle size of the KBr and the sample powders (after grinding with ShakIR) are in the range of 0.7-2.2 μm, examined using a Leica DM2700P microscope. The scattering effect caused by the fine particles is minimal in the longer wavelengths and grows a little (a few percent of the maximum transmittance) towards the short wavelengths because the particle size is comparable to the short wavelengths. However, using the transmittance of a pure KBr pellet as a reference can correct for light scattering losses in the pellet. The resulting spectra (Fig. 2) can reflect the true features from the sample itself. In addition, the transmittance of the film samples of the 300 K and 400 K haze analogue (see Figure S2) is consistent with that of the pellet samples, confirming that the potential scattering effect in the pellet has been largely corrected. Note that there is small difference in the spectral slope in the visible (0.4-0.625 μm) between the pellet and the film, which could suggest the film is more absorbing in this region or could be caused by the light loss at the interfaces in the film system. However, we have done the transmittance measurements for the film on $CaF_2$ substrate and the powder sample of our standard Titan tholin (produced with PHAZER using 5% $CH_4$ in $N_2$) and the spectra in the MIR (2.5-10 μm) shows that the spectral features for the organic functional groups are identical.

*4.4 Optical Constants Derivation*

With the data from the measurements described above, we can derive the optical constants (complex refractive index, n+ik, where n is the real refractive index and k is the imaginary part of the refractive index) of the haze analogues. Based on the Beer-Lambert law, the relation between transmittance ($T$) and the absorption coefficient ($\alpha$) for non-scattering samples can be expressed as Equation 1 (Eq. 1):

$$-lnT = -ln\frac{I}{I_0} = \alpha(\nu) * d \qquad \text{Eq. 1}$$

where $T$ is the transmittance, $I$ is the light intensity passing through the sample (haze analogue/KBr pellet), $I_0$ is the light intensity of the reference (pure KBr pellet), $\alpha$ is the absorption coefficient, $\nu$ is the wavenumber, and $d$ is the effective thickness of the sample. The effective thickness ($d$) of the sample in the pellet is equal to:

$$d = \frac{m}{\pi r^2 * \rho} \qquad \text{Eq. 2}$$

in which $m$ is the mass of the haze analogues in the pellet, $r$ is the pellet radius (6.5 mm), and $\rho$ is the measured density of the haze analogues.

The absorption coefficient ($\alpha$) can also be expressed as:

$$\alpha(\nu) = 4\pi k \nu \qquad \text{Eq. 3}$$

Then, the imaginary part of the refractive index ($k$) can be determined:

$$k(\nu) = \frac{\alpha(\nu)}{4\pi\nu} = \frac{1}{4\pi\nu d} ln\frac{I_0}{I} \qquad \text{Eq. 4}$$

For each sample, we obtain three sets of $k$ values independently by measuring the transmittance of three pellets with different effective thicknesses. The final $k$ values are determined by averaging the three sets of k values.

In order to calculate the real refractive index ($n$) based on the subtractive Kramers-Kronig (SKK) relation[63,72,73] between $n$ and $k$, we need to know an anchor point $n_0$ value. From the interference fringes on the reflectance spectra of the film samples at two angles (15° and 45°) of incidence (see Extended Data Fig. 1), we determine $n_0$ values from 0.4 to 1.1 μm using Eq. 5.[74]

$$n_0 = \sqrt{\frac{sin^2\theta_1*(\Delta\nu_1)^2 - sin^2\theta_2*(\Delta\nu_2)^2}{(\Delta\nu_1)^2 - (\Delta\nu_2)^2}} \qquad \text{Eq. 5}$$

where $\theta_1$ and $\theta_2$ are the two different angles of incidence in our reflectance measurements of the film samples (15° and 45°), and $\Delta v_1$ and $\Delta v_2$ are the average fringe spacing in the reflectance spectrum at respective incidence angle (see Table S4).

With the determined imaginary part of the refractive index ($k$) and an anchor point $n_0$ value, we can calculate the real refractive index ($n$) based on the subtractive Kramers-Kronig (SKK) relation[63,72,73] between $n$ and $k$ as in Eq. 6:

$$n(v) = n_0 + \frac{2(v^2 - v_0^2)}{\pi} P \int_0^\infty \frac{v'k(v')}{(v'^2 - v^2)(v'^2 - v_0^2)} dv' \qquad \text{Eq. 6}$$

where $v$ is wavenumber (cm$^{-1}$), $n_0$ is the real refraction index at $v_0$, and $P$ indicates the Cauchy principal value. The principal $P$ value is integrated for the entire wavelength range; we use the $k$ values calculated (Eq. 4) from 350 to 25000 cm$^{-1}$ in the integrand and assume constant $k$ values for wavelengths beyond the measured range. The assumption is generally valid unless there are large local absorption peaks outside the measurement range. Using an anchor point ($n_0$) can reduce the uncertainty for the numerical integration as demonstrated in previous studies.[63,73] The $n_0$ value determined at 0.625 μm ($v_0$=16000 cm$^{-1}$, $n_0$=1.6213 for 400 K sample and, $n_0$=1.6027 for 300 K sample) from Eq. 5 is used as the anchor point for the numerical integration in Eq. 6 to calculate the real refractive index ($n$) from 0.4 to 28.6 μm using the SKK relation. We used different anchor points ($n_0$ at different wavelengths) for the calculation and found that the yielded $n$ values (see Figure S3) are very close (the relative standard deviation is less than 2%). In the sensitivity analysis (Figure S3), we derived n using the $n_0$ anchor only for wavelengths above 0.625 μm to make sure there is no risk of using an anchor with a value of n that would be different from powder to film.

*4.5 Atmospheric Transmission Spectra Simulation*

We use the open-source aerosol modeling code *Virga*[48] and the open-source radiative transfer suite *PICASO*[49] to compute haze Mie properties and implement them into a synthetic transmission spectrum model of an exoplanetary atmosphere. To generate hazy atmospheric model spectra, we require a stellar radius, planetary mass and radius, pressure-temperature profile, an atmospheric gas mixture, and haze profiles.

For the atmospheric composition of our synthetic spectra, we input the system parameters of GJ 1214 b for the stellar and planetary mass and radius[44] and the same gas mixing ratios as in the experimental run of the 400 K experiment shown in Fig. 1 into *PICASO*, using a simple parametrized isothermal pressure-temperature profile at GJ 1214 b's $T_{eq}$ of 600 K. Into *Virga*, we input the optical properties of our experimental haze analogues as their real and imaginary refractive indices *n* and *k*, binned down from the native measured resolution to a spacing of 10 nm, which we verify is still a fine enough sampling to capture the spectral features of the haze. We then used *Virga*'s optics functionality using PyMieScatt[75] to generate wavelength-dependent Mie coefficients using haze particle size distributions as measured in the laboratory for the 400 K haze analogue by He et al. 2018[76]. Note that the use of Mie scattering is an approximation, because the particle morphology in an actual atmosphere is more complex than that of our laboratory samples and would impact the spectra differently than the spherical particles in Virga's scattering model. In addition to the optical properties and particle size of the experimental haze analogues, we also need the vertical extent and mass loading of the haze layer within *Virga* in order to compute a synthetic atmospheric transmission spectrum with *PICASO*. *Virga* currently does not have the capability to self-consistently compute photochemical haze profiles. Therefore, for simplicity, we implement a homogeneous haze layer from a pressure of 0.1 bar to 0.1 μbar with haze mass loading of ~25 particle/cm$^3$. These values are consistent with the extent of haze in Titan's mesosphere[77] as well as self-consistent haze formation modeling that has been performed for exoplanets.[78] The resulting optical depth of the wavelength-dependent haze layers as a function of pressure and wavelength from our model set-up is shown in Extended Data Fig. 2. For the Titan haze analogue cases, we use the same settings as described above for haze layers modeled with our 400K haze analogue, but substitute the refractive indices of Khare et al. (1984)[14]. Similarly, we also model the spectra (Extended Data Fig. 4) using the optical constants reported by Corrales et al. (2023).[56]

The purpose of our spectra simulation is to show the effect of the different haze optical properties on the atmospheric transmission spectra rather than perform a full model parameter study. Simplified assumptions were used in our simulation, such as atmospheric pressure-temperature profiles, the particle radius and mass distribution. Future model development and modeling studies, employing self-consistent pressure-temperature profiles including the radiative effects of the haze[17], and full microphysics treatment of particle parameters[79-83], are required to fully explore haze impacts on observations of exoplanet atmospheres.

**Data availability**

The data resulting from this study are provided in the Article and Supplementary Information. Data associated with Figure 3 and Figure 4 are available in the Johns Hopkins University Data Archive with the following https://doi.org/10.7281/T1/NEACHP. Other data that support the plots within this paper and other findings of this study are available from the corresponding author upon reasonable request.

**References**

1. Kreidberg, L. et al. Clouds in the atmosphere of the super-Earth exoplanet GJ1214b. *Nature* **505**, 69–72 (2014).
2. Knutson, H. A., Benneke, B., Deming, D. & Homeier, D. A featureless transmission spectrum for the Neptune-mass exoplanet GJ436b. *Nature* **505**, 66–68 (2014).
3. Knutson, H. A. et al. Hubble space telescope near-IR transmission spectroscopy of the super-Earth HD 97658b. *Astrophys. J.* **794**, 155 (2014).
4. May, E. M., Gardner, T., Rauscher, E., & Monnier, J. D. MOPSS. II. Extreme Optical Scattering Slope for the Inflated Super-Neptune HATS-8b. *Astron. J.* **159**(1), 7 (2019).
5. Dragomir, D. et al. Rayleigh scattering in the atmosphere of the warm exo-Neptune GJ 3470b. *Astrophys. J.* **814**, 102 (2015).
6. JWST Transiting Exoplanet Community Early Release Science Team. Identification of carbon dioxide in an exoplanet atmosphere. *Nature* (2022). https://doi.org/10.1038/s41586-022-05269-w

7. Gao, P. et al. Aerosol composition of hot giant exoplanets dominated by silicates and hydrocarbon hazes. *Nat. Astron*. **4**, 951–956 (2020).
8. Morley, C. V., Fortney, J. J., Kempton, E. M. R., Marley, M. S., Visscher, C., & Zahnle, K. Quantitatively assessing the role of clouds in the transmission spectrum of GJ 1214b. *Astrophys. J.* **775**(1), 33 (2013).
9. He, C. et al. Photochemical haze formation in the atmospheres of super- Earths and mini-Neptunes. *Astron. J.* **156**, 38 (2018).
10. Hörst, S. M. et al. Haze production rates in super-Earth and mini-Neptune atmosphere experiments. *Nat. Astron.* **2**, 303–306 (2018).
11. He, C., Hörst, S. M., Lewis, N. K., et al. Sulfur-driven haze formation in warm $CO_2$-rich exoplanet atmospheres. *Nat. Astron.* **4**, 986-993 (2020).
12. He, C., Hörst, S. M., Lewis, N. K., et al. Haze formation in warm $H_2$-rich exoplanet atmospheres. *Planet. Sci. J.* **1**, 51 (2020).
13. Gao, P., Wakeford, H. R., Moran, S. E., & Parmentier, V. Aerosols in Exoplanet Atmospheres. *JGR: Planets* **126**(4), e06655, (2021).
14. Khare, B.N., et al. Optical constants of organic tholins produced in a simulated Titanian atmosphere: From soft X-ray to microwave frequencies. *Icarus* **60**, 127–137 (1984).
15. Chang, H. & Charalampopoulos, T. T. Determination of the Wavelength Dependence of Refractive Indices of Flame Soot. *In Proceedings of the Royal Society of London. Series A* **430**, 577–591 (1990).
16. Lavvas, P., & Koskinen, T. Aerosol properties of the atmospheres of extrasolar giant planets. *Astrophys. J.* **847**(1), 32 (2017).
17. Morley, C. V., Fortney, J. J., Marley, M. S., Zahnle, K., Line, M., Kempton, E., et al. Thermal emission and reflected light spectra of super Earths with flat transmission spectra. *Astrophys. J.* **815**(2), 110 (2015).
18. Moran, S. E. et al. Chemistry of temperate super-Earth and mini-Neptune atmospheric hazes from laboratory experiments. *Planet. Sci. J.* **1**, 17 (2020).
19. Tsiaras, A., Waldmann, I. P., Tinetti, G., Tennyson, J. & Yurchenko, S. N. Water vapour in the atmosphere of the habitable-zone eight-Earth-mass planet K2-18b. *Nat. Astron.* **3**, 1086–1091 (2019).
20. Benneke, B., Wong, I., Piaulet, C., Knutson, H. A., Lothringer, J., Morley, C. V., et al. Water vapor and clouds on the habitable-zone sub-Neptune exoplanet K2-18b. *Astrophys. J. Lett.* **887**(1), L14 (2019).
21. Mulders, G., Ciesla, F., Min, M. & Pascucci, I. The snow line in viscous disks around low-mass stars: implications for water delivery to terrestrial planets in the habitable zone. *Astrophys. J.* **807**, 9–15 (2015).
22. Kite, E. S. & Ford, E. B. Habitability of exoplanet waterworlds. *Astrophys. J.* **864**, 75–102 (2018).

23. Zeng, L. et al. Growth model interpretation of planet size distribution. *Proc. Natl Acad. Sci. USA* **116**, 9723–9728 (2019).
24. Kite, E. S. & Schaefer, L. Water on hot rocky exoplanets. *Astrophys. J. Lett.* **909**, L22 (2021).
25. Luque, R., & Pallé, E. Density, not radius, separates rocky and water-rich small planets orbiting M dwarf stars. *Science* **377**, 1211-1214 (2022).
26. Chachan, Y., Jontof-Hutter, D., Knutson, H. A., Adams, D., Gao, P., Benneke, B., et al. A featureless infrared transmission spectrum for the super-puff planet Kepler-79d. *Astrophys. J.* **160**(5), 201 (2020).
27. Libby-Roberts, J. E., Berta-Thompson, Z. K., Désert, J.-M., Masuda, K., Morley, C. V., Lopez, E. D., et al. The featureless transmission spectra of two super-puff planets. *Astron. J.* **159**(2), 57 (2020).
28. Adams, D., Gao, P., de Pater, I. & Morley, C. V. Aggregate hazes in exoplanet atmospheres. *Astrophys. J.* **874**, 61 (2019).
29. Gao, P., & Zhang, X. Deflating Super-puffs: Impact of photochemical hazes on the observed mass-radius relationship of low-mass planets. *Astrophys. J.* **890**(2), 93 (2020).
30. Ohno, K., & Tanaka, Y. A. Grain growth in escaping atmospheres: Implications for the radius inflation of super-puffs. *Astrophys. J.* **920**(2), 124 (2021).
31. He, C. et al. Carbon monoxide affecting planetary atmospheric chemistry. *Astrophys. J. Lett.* **841**, L31 (2017).
32. Moran, S. E. et al. Triton haze analogs: The role of carbon monoxide in haze formation. *JGR-Planets* **127**, e2021JE006984 (2022).
33. Rao, C. N. R. Ultra-violet and visible spectroscopy: chemical applications, Butterworths, London, pp. 242 (1975).
34. van Krevelen D. W. & te Nijenhuis K. Chapter 10: Optical Properties, in Properties of Polymers, Elsevier, Oxford, UK, pp. 287-320 (2009).
35. Lin-Vien, D., Colthup, N. B., Fateley, W. G. & Grasselli, J. G. The Hand-book of Infrared and Raman Characteristic Frequencies of Organic Molecules. Academic Press, San Diego. 503 p (1991).
36. Socrates, G. Infrared and Raman Characteristic Group Frequencies. Wiley, Chichester. 347 p (2001).
37. Duvernay, F., et al. Carbodiimide production from cyanamide by UV irradiation and thermal reaction on amorphous water ice. *J Phys Chem A.* **109**(4), 603-608 (2005).
38. Khare, B. N., et al. Analysis of the time-dependent chemical evolution of Titan haze tholin. *Icarus* **160**(1), 172-182 (2002).
39. Imanaka, H., et al. Laboratory experiments of Titan tholin formed in cold plasma at various pressures: implications for nitrogen-containing polycyclic aromatic compounds in Titan haze. *Icarus* **168**(2), 344-366 (2004).

40. Vinatier, S., et al. Optical constants of Titan's stratospheric aerosols in the 70–1500 cm$^{-1}$ spectral range constrained by Cassini/CIRS observations. *Icarus* **219**(1), 5-12 (2012).

41. Ohno, K., Zhang, X., Tazaki, R., & Okuzumi, S. Haze formation on Triton. *Astrophys. J.* **912**(1), 37 (2021).

42. Zhang, X., Strobel, D. F., & Imanaka, H. Haze heats Pluto's atmosphere yet explains its cold temperature. *Nature* **551**(7680), 352-355 (2017).

43. Arney, G. N., et al. Pale orange dots: the impact of organic haze on the habitability and detectability of Earthlike exoplanets. *Astrophys. J.* **836**(1), 49 (2017).

44. Cloutier, R., et al. A More Precise Mass for GJ 1214 b and the Frequency of Multiplanet Systems Around Mid-M Dwarfs. *Astron. J.* **162**, 174 (2021).

45. Lora, J.M., et al. Atmospheric Circulation, Chemistry, and Infrared Spectra of Titan-like Exoplanets around Different Stellar Types. *Astrophys. J.* **853**, 58 (2018).

46. Teal, D.J., et al. Effects of UV Stellar Spectral Uncertainty on the Chemistry of Terrestrial Atmospheres. *Astrophys. J.* **927**, 90. (2022).

47. Ackerman, A.S. & Marley, M.S. Precipitating Condensation Clouds in Substellar Atmospheres. *Astrophys. J.* **556**, 872-884 (2001).

48. Rooney, C.M., et al. A New Sedimentation Model for Greater Cloud Diversity in Giant Exoplanets and Brown Dwarfs. *Astrophys. J.* **925**, 33 (2022).

49. Batalha, N.E., et al. Exoplanet Reflected-light Spectroscopy with PICASO. *Astrophys. J.* **878**, 70 (2019).

50. Ohno, K. & Kawashima, Y. Super-Rayleigh Slopes in Transmission Spectra of Exoplanets Generated by Photochemical Haze. *Astrophys. J. Lett.* **895**, L47 (2020).

51. Ahrer, E.-M., et al., Atmospheric water and chemistry in the exoplanet WASP-39b with JWST NIRCam. *Nature*, in press (2022).

52. Alderson, L., et al., The molecular inventory of the exoplanet WASP-39b with JWST NIRSpec G395H. *Nature*, in press (2022).

53. Feinstein, A.D., et al. Atmospheric composition and clouds in the exoplanet WASP-39b with JWST NIRISS. *Nature*, in press (2022).

54. Rustamkulov, Z., et al., A panchromatic spectrum of the exoplanet WASP-39b with JWST NIRSpec PRISM. *Nature*, in press (2022).

55. Sing, D.K., et al., A continuum from clear to cloud hot-Jupiter exoplanets without primordial water depletion. *Nature* **529**, 7584 (2016).

56. Corrales, L. et al. Photochemical hazes can trace the C/O ratio in exoplanet atmospheres. arXiv preprint arXiv:2301.01093 (2023).

57. Lupu, R. E., Marley, M. S., Lewis, N., Line, M., Traub, W. A., & Zahnle, K. Developing atmospheric retrieval methods for direct imaging spectroscopy of gas giants in reflected light. I. Methane abundances and basic cloud properties. *Astron. J.* **152**(6), 217 (2016).

58. Lacy, B., Shlivko, D., & Burrows, A. Characterization of exoplanet atmospheres with the optical coronagraph on WFIRST. *Astron. J.* **157**(3), 132 (2019).
59. Steinrueck, M. E., et al. 3D simulations of photochemical hazes in the atmosphere of hot Jupiter HD 189733b, *Monthly Notices of the Royal Astronomical Society* **504**(2), 2783–2799 (2021).
60. Moses, J. I. et al. Compositional diversity in the atmospheres of hot Neptunes, with application to GJ 436b. *Astrophys. J.* **777**, 34 (2013).
61. Ramirez, S. I., Coll, P., da Silva, A., et al. Complex refractive index of Titan's aerosol analogues in the 200–900 nm domain. *Icarus* **156**(2), 515-529 (2002).
62. Tran, B. N., Joseph, J. C., Ferris, J. P., et al. Simulation of Titan haze formation using a photochemical flow reactor: The optical constants of the polymer. *Icarus* **165**(2), 379-390 (2003).
63. Vuitton, V., Tran, B. N., Persans, P. D. & Ferris, J. P. Determination of the complex refractive indices of Titan haze analogs using photothermal deflection spectroscopy. *Icarus* **203**(2), 663-671(2009).
64. Imanaka, H., Cruikshank, D. P., Khare, B. N., & McKay, C. P. Optical constants of Titan tholins at mid-infrared wavelengths (2.5–25 μm) and the possible chemical nature of Titan's haze particles. *Icarus* **218**(1), 247-261 (2012).
65. Sciamma-O'Brien, E., Dahoo, P. R., Hadamcik, E., Carrasco, N., Quirico, E., Szopa, C., & Cernogora, G. Optical constants from 370 nm to 900 nm of Titan tholins produced in a low pressure RF plasma discharge. *Icarus* **218**(1), 356-363 (2012).
66. Mahjoub, A., Carrasco, N., Dahoo, P. R., Gautier, T., Szopa, C., & Cernogora, G. Influence of methane concentration on the optical indices of Titan's aerosols analogues. *Icarus* **221**(2), 670-677 (2012).
67. Brassé, C., Muñoz, O., Coll, P., & Raulin, F. Optical constants of Titan aerosols and their tholins analogs: Experimental results and modeling/observational data. *Planetary and Space Science* **109**, 159-174 (2015).
68. Gavilan, L., Carrasco, N., Hoffmann, S. V., Jones, N. C., & Mason, N. J. Organic Aerosols in Anoxic and Oxic Atmospheres of Earth-like Exoplanets: VUV-MIR Spectroscopy of CHON Tholins. *Astrophys. J.* **861**(2), 110 (2018).
69. Jovanović, L. et al. Optical constants of Pluto aerosol analogues from UV to near-IR. *Icarus* **362**, 114398 (2021).
70. He, C., Hörst, S. M., Radke, M., & Yant, M. Optical Constants of a Titan Haze Analog from 0.4 to 3.5 μm Determined Using Vacuum Spectroscopy. *The Planetary Science Journal* **3**(1), 25 (2022).
71. Myers, T. L., et al. Obtaining the complex optical constants n and k via quantitative absorption measurements in KBr pellets. *Chemical, Biological, Radiological, Nuclear, and Explosives (CBRNE) Sensing XX*. 11010. SPIE (2019).

72. Wood, B. E., & Roux, J. A. Infrared optical properties of thin $H_2O$, $NH_3$, and $CO_2$ cryofilms. *JOSA,* **72**(6), 720-728 (1982).
73. Toon, O. B., et al. Infrared optical constants of $H_2O$ ice, amorphous nitric acid solutions, and nitric acid hydrates. *JGR-Atmospheres* **99**(D12), 25631-25654 (1994).
74. Padera, F. Measuring Absorptance (k) and Refractive Index (n) of Thin Films with the PerkinElmer LAMBDA 1050+ High Performance UV/Vis/NIR Spectrometers. PerkinElmer Application Note.
75. Sumlin, B. J., Heinson, W. R., & Chakrabarty, R. K. Retrieving the aerosol complex refractive index using PyMieScatt: A Mie computational package with visualization capabilities. *Journal of Quantitative Spectroscopy and Radiative Transfer* **205**, 127-134, (2018).
76. He, C. Hörst, S.M., Lewis, N.K., et al. Laboratory Simulations of Haze Formation in the Atmospheres of Super-Earths and Mini-Neptunes: Particle Color and Size Distribution. *Astrophys. J. Lett.* **856,** 1 (L3). 2018.
77. Lavvas, P., Yelle, R.V., and Vuitton, V. The detached haze layer in Titan's mesosphere. *Icarus* **201** (2), 626-633 (2009).
78. Kawashima, Y. & Ikoma, M. Theoretical Transmission Spectra of Exoplanet Atmospheres with Hydrocarbon Haze: Effect of Creation, Growth, and Settling of Haze Particles. I. Model Description and First Results. *Astrophys. J.* **853**, 7 (2018).
79. Trainer, M.G., et al. The Influence of Benzene as a Trace Reactant in Titan Aerosol Analogs. *Astrophys. J. Lett.* **766,** L4 (2013).
80. Lavvas, P., et al. Aerosol growth in Titan's Ionosphere. *Proc. Natl Acad. Sci. USA* **110**, 8 (2013).
81. Yoon, Y.H., et al., The Role of Benzene Photolysis in Titan Haze Formation. *Icarus* **233**, 233-241 (2014).
82. Gao, P. & Benneke, B. Microphysics of KCl and ZnS Clouds on GJ 1214 b. *Astrophys. J.* **863**, 165 (2018).
83. Ohno, K. & Okuzumi, S. A Condensation-Coalescence Cloud Model for Exoplanetary Atmospheres: Formulation and Test Applications to Terrestrial and Jovian Clouds. *Astrophys. J.* **835**, 261 (2017).

**Acknowledgements**

This work was supported by the NASA Astrophysics Research and Analysis Program NNX17AI87G and the NASA Exoplanets Research Program 80NSSC20K0271.

**Author contributions**

C.H., M.R., S.E.M., S.M.H., N.K.L., M.S.M., and J.I.M. conceived the study. J.I.M. calculated the starting gas mixtures. C.H. carried out the experiments. C.H. and M.R. performed the optical measurements. S.E.M. and C.H. simulated the synthetic transmission spectra. C.H. conducted the data analysis and prepared the manuscript. All authors participated in discussions regarding interpretation of the results and edited the manuscript.

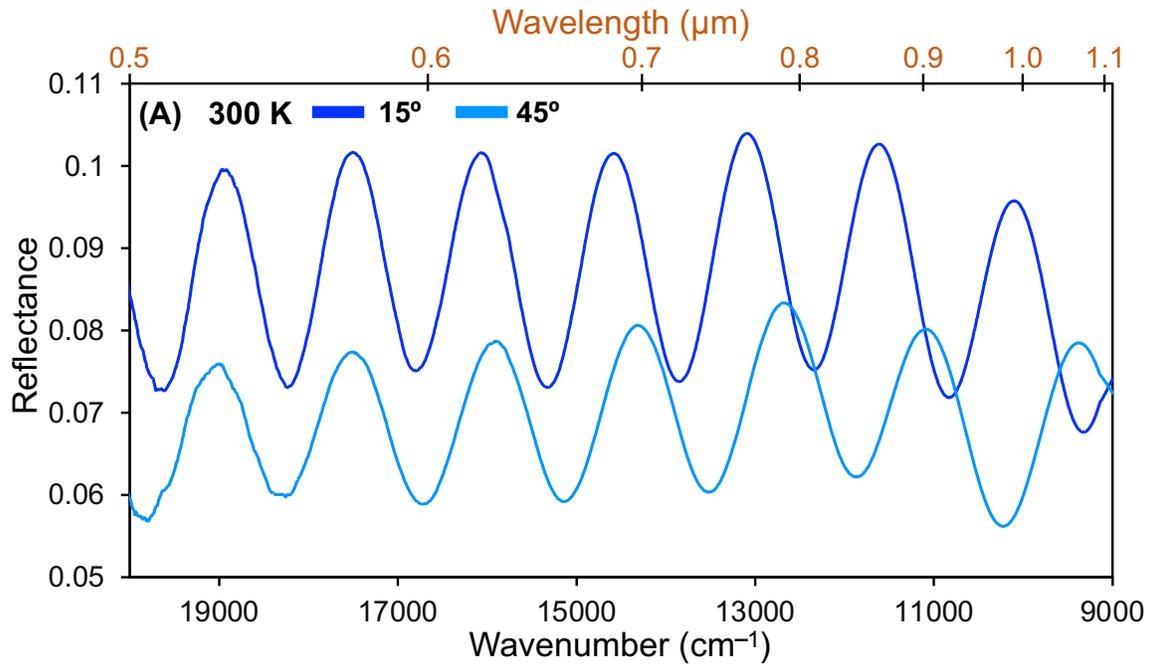

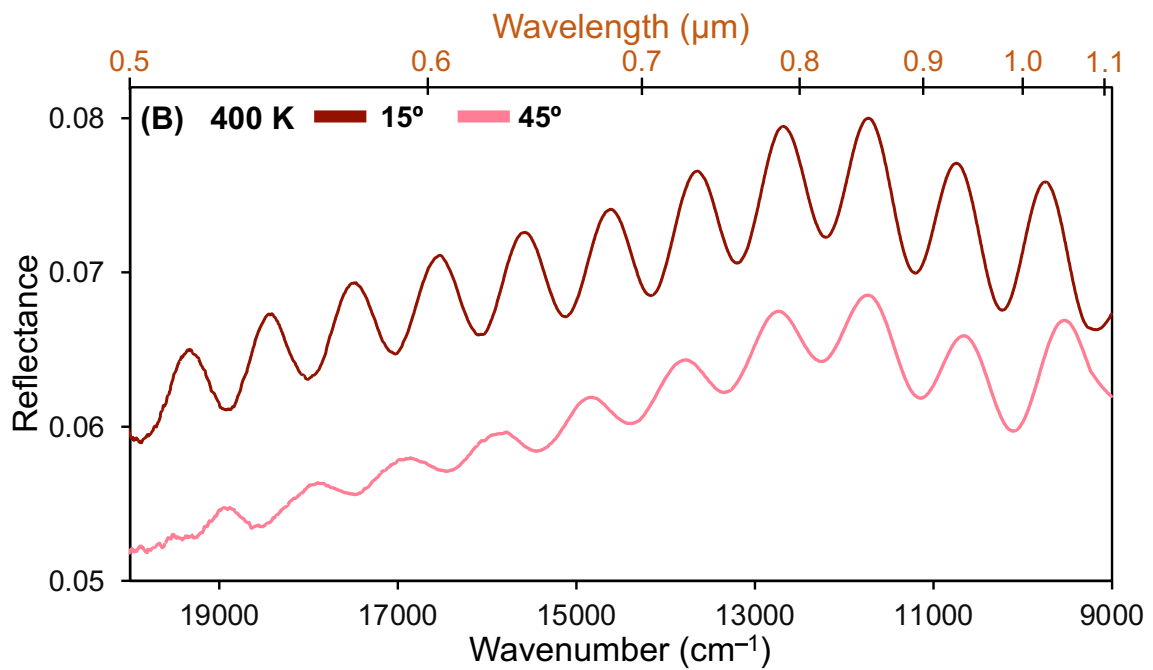

**Extended Data Fig. 1**. Reflectance spectra of two exoplanet haze analogues formed in water-rich gas mixtures at 300 K (A) and 400 K (B). Using Seagull Variable Angle Reflection Accessory, we measure the reflectance of each haze analogue film at two different angles of incidence, 15° (darker shade) and 45° (lighter shade). The spectra from 20000 to 9000 cm$^{-1}$ (0.5 to 1.1 μm) are shown here. From the interference fringes on the reflectance spectra at two different angles, the $n_0$ values of the samples at corresponding wavelengths can be determined using Eq. 5.

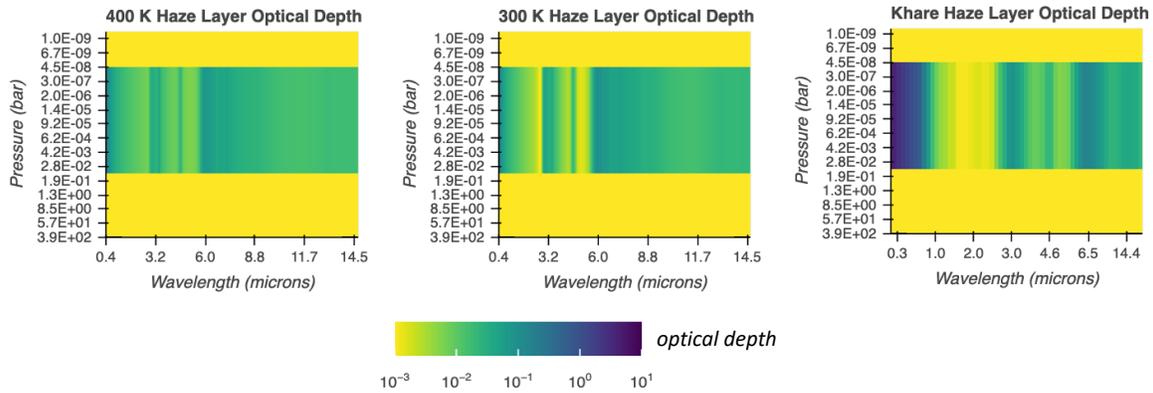

**Extended Data Fig. 2**. Modeled haze slab profiles as a function of pressure and wavelength as implemented with the *Virga*-derived Mie coefficients. The color bar indicates the optical depth at each wavelength, with darker shading corresponding to higher optical depths. The haze layer extends from 0.1 bar to 0.1 μbar with a haze particle radii distribution centered around 25 nm. Note that the Khare et al. (1984) data has wider wavelength coverage than shown, but at lower resolution than our measured optical properties.

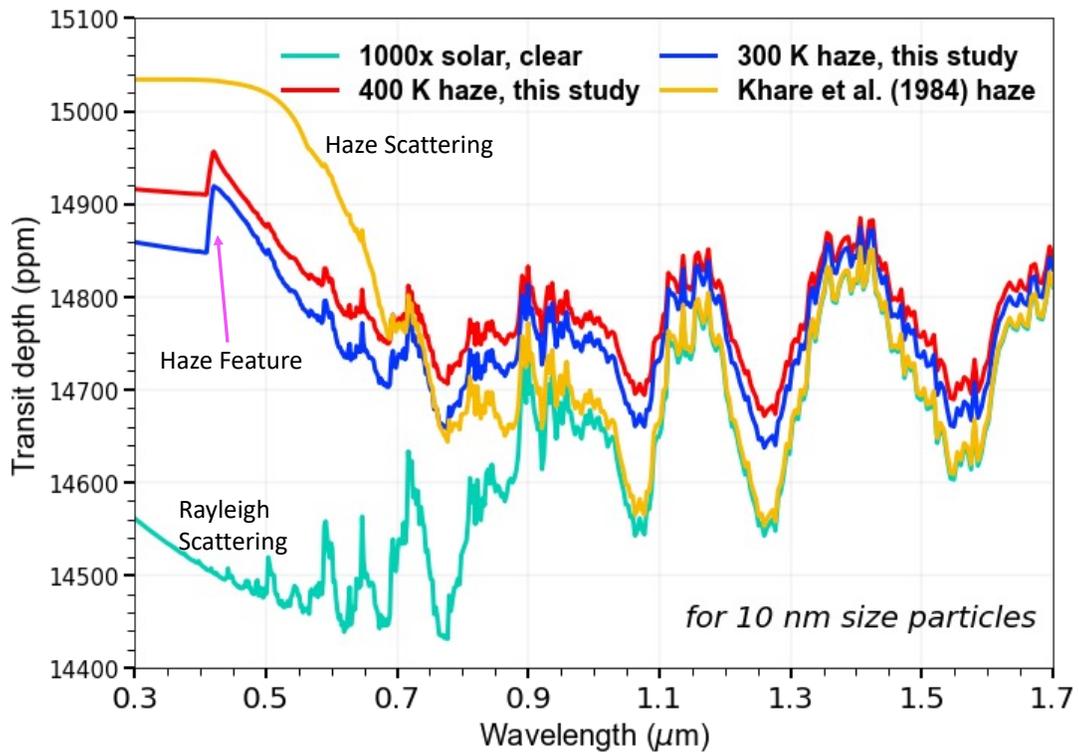

**Extended Data Fig. 3.** Model spectra of a water-rich atmosphere around a GJ 1214 b -like planet. We show the effect of our newly measured haze optical properties using small radii (10 nm) haze particles, focusing on the wavelength range accessible to *Hubble*. The method and settings for generating the spectra here are the same as described in 4.5, except the haze particle radii (10 nm) and haze mass loading (4 particles/cm$^3$). With sufficiently small particles, the large scattering slopes between different haze compositions are differentiable with *Hubble*'s ultraviolet/visible capabilities even if such hazes less strongly impact the NIR.

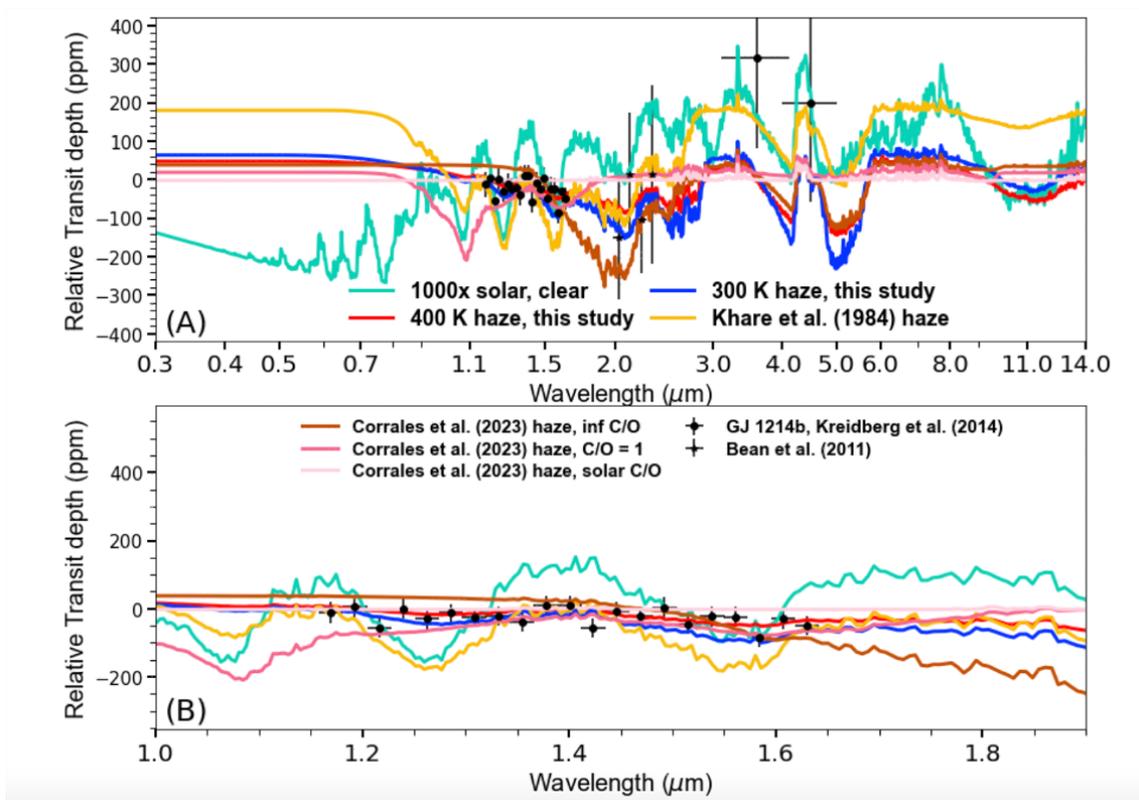

**Extended Data Fig. 4.** Model spectra of a water-rich atmosphere around a GJ 1214 b -like planet. Here we add the synthetic spectra using the optical constants reported by Corrales et al. (2023)[56] to show the effect of different haze optical properties. The method and settings for generating the spectra here are the same as described in 4.5. Compared to our water-rich atmospheric haze analogues and Khare Titan haze analogue, the haze analogues from Corrales et al. (2023)[56] more heavily mute the atmospheric spectral features in the longer wavelengths (2.5 to 14 μm). In the short wavelengths, the haze analogues from Corrales et al. (2023)[56] mute the atmospheric features at similar level as our water-rich atmospheric haze analogues.